\documentclass[prd,aps,preprint,tightenlines]{revtex4b3}
\usepackage{epsfig}
\usepackage{multicol}

\begin{document}

\newcommand{\TeV}{\,{\rm TeV}}
\newcommand{\GeV}{\,{\rm GeV}}
\newcommand{\MeV}{\,{\rm MeV}}
\newcommand{\keV}{\,{\rm keV}}
\newcommand{\eV}{\,{\rm eV}}
\def\ap{\approx}
\newcommand{\bea}{\begin{eqnarray}}
\newcommand{\eea}{\end{eqnarray}}
\def\beq{\begin{equation}}
\def\eeq{\end{equation}}
\def\haf{\frac{1}{2}}
\def\lpp{\lambda''}
\def\ccg{\cal G}
\def\slash#1{#1\!\!\!\!\!\!/}
\def\u{{\cal U}}

%%%%%%%%%%%%%%%%%%%%% Main Text %%%%%%%%%%%%%%%%
\setcounter{page}{1}
%\renewcommend{\arraystretch}{1.3}
\preprint{KAIST-TH 01/01, SNUTP 01-002, hep-ph/0101026}

\title{Axino as a sterile neutrino and $R$ parity violation}
 
\author{Kiwoon Choi$^a$, Eung Jin Chun$^b$
and Kyuwan Hwang$^a$}

\affiliation{$^a$Korea Advanced Institute of Science and Technology,
        Taejon 305-701, Korea  \\
$^b$Department of Physics, Seoul National University, Seoul 151-747, Korea}

%\date{\today}

\begin{abstract}
We point out that axino
can be a natural candidate for sterile neutrino
which would accommodate the LSND data
with atmospheric and solar neutrino oscillations.
It is shown that the so called $3+1$ scheme can be
easily realized
when supersymmetry breaking is mediated by gauge interactions
and also $R$-parity is properly broken.
Among the currently possible solutions to the
solar neutrino problem, only the large angle MSW 
oscillation is allowed  in this scheme.
The weak scale value of the Higgs $\mu$ parameter 
and the required size of $R$-parity violation
can be understood by means of
spontaneously broken Peccei-Quinn symmetry.
\end{abstract}


\maketitle

%\begin{multicols}{2}

Current data from the atmospheric and 
solar neutrino experiments are beautifully explained  
by oscillations among three active neutrino species \cite{glb}. 
Another data in favor of neutrino oscillation has been obtained 
in the LSND experiment \cite{lsnd}.
Reconciliation of these experimental results 
requires three distinct mass-squared differences, 
implying the existence of a sterile neutrino $\nu_s$ \cite{moha}.
In the four-neutrino oscillation framework, there are two 
possible scenarios: 
the $2+2$ scheme in which  two pairs of close mass eigenstates are separated
by the LSND mass gap $\sim 1$ eV
and the $3+1$ scheme in which one mass is isolated from the other
three by the LSND mass gap. 
It has been claimed  that the LSND results can  
be compatible with various short-baseline 
experiments only in the context of the 2+2 scheme \cite{2by2}.  
However, according to the new LSND results \cite{lsnd},
the allowed parameter regions are shifted to smaller mixing angle,
accepting the $3+1$ scheme \cite{smirnov,barger,giunti}.
Although it can be realized in a rather limited parameter space,
the 3+1 scheme is attractive since the fourth (sterile) neutrino 
can be added without changing the most favorable picture that 
the atmospheric and solar neutrino data are explained
by   the predominant $\nu_\mu \to \nu_\tau$ 
and $\nu_e \to \nu_\mu, \nu_\tau$ oscillations, respectively.
In particular,  the $3+1$ scheme with the heaviest $\nu_s$ 
would be an interesting explanation of all existing neutrino data.
On the other hand, it is rather difficult to find a well-motivated 
particle physics model which would yield the desired four-neutrino masses and
mixing in a consistent way \cite{yanagida}.

In this paper, we show that the 3+1 scheme can be easily realized
in supersymmetric model with $U(1)$ Peccei-Quinn (PQ) symmetry
when supersymmetry (SUSY) breaking is mediated by gauge interactions.
In this model, axino can be as light as 1 eV, so plays the role of sterile
neutrino \cite{chun}.
A proper axino-neutrino mixing 
can be  induced by $R$-parity violating couplings which 
appear as a consequence of spontaneously broken $U(1)_{PQ}$.
It turns out that only the large angle MSW solution to the solar neutrino
problem is allowed in our model.
The weak scale value of the Higgs $\mu$-parameter and the
required size of $R$-parity violation can be understood
by means of the Frogatt-Nielsen mechanism \cite{nielson} of
spontaneously broken $U(1)_{PQ}$.

The model under consideration includes
three sectors: the observable sector, the SUSY-breaking sector,
and the PQ sector.  The first
contains the usual quarks, leptons,
and two Higgs superfields, {\it i.e.} the MSSM superfields.
The second contains a gauge-singlet 
Goldstino superfield
$X$ and the gauge-charged messenger superfields $Y,Y^c$.
Finally the third contains gauge-singlet superfields $S_k$ 
which break $U(1)_{PQ}$ by their vacuum expectation values (VEV), 
as well as gauge-charged superfields $T,T^c$
which have  the Yukawa coupling with some of $S_k$.
The superpotential of the model is given by
\beq
W=h S_3(S_1S_2-f_{PQ}^2)+\kappa S_1TT^c+
\lambda XYY^c + W_{\rm MSSM}+W_{\rm SB}
\eeq
where $W_{\rm MSSM}$ involves the MSSM fields, and
$W_{\rm SB}$ describes SUSY breaking dynamics
enforcing $X$ develop 
 a SUSY breaking VEV:
$\lambda X =M_X+\theta^2 F_X$.
This VEV generates soft masses of the MSSM fields
\cite{gmsb},
 $m_{\rm soft}=\frac{\alpha}{2\pi}F_X/M_X$, which are assumed
to be of order the weak scale.
One can easily arrange the symmetries of the model,
$U(1)_{PQ}$ and an additional discrete symmetry to get
\bea
W_{\rm MSSM}&&=y^{E}_{ij}H_1L_iE^c_j+y^{D}_{ij}H_1Q_iD^c_j+
y^{U}_{ij}H_2Q_iU^c_j 
 +\frac{y_0}{M_*}S_1^2H_1H_2\nonumber\\
&&+\frac{y^{\prime}_i}{M_*^2}S_1^3L_iH_2 
+\frac{\gamma_{ijk}}{M_*}S_1L_iL_jE^c_k
+\frac{\gamma^{\prime}_{ijk}}{M_*}S_1L_iQ_jD^c_k, 
\eea
where the Higgs, quark and lepton superfields are in obvious
notations. Here $M_*$ denotes the fundamental scale of the model which will
be taken to be the grand unification scale, $M_{GUT}\approx 10^{16}$ GeV.
Full details of the model will be presented elsewhere
\cite{cch}.

Let $A=(\phi+ia)+\theta\tilde{a}+\theta^2 F_A$ denote the axion
superfield where $a$, $\phi$ and $\tilde{a}$ are
the axion, saxion and axino, respectively.
It is then convenient
to parameterize $S_1$ and $S_2$ as 
$S_1=Se^{A/f_{PQ}}$ and  $S_2=Se^{-A/f_{PQ}}$.
Note that the VEV of $S$ is uniquely determined to be
$\langle S\rangle=f_{PQ}$ which would correspond
to the axion decay constant {\it if} the saxion is stabilized at
$e^{\phi/f_{PQ}}\approx 1$.
After integrating out the SUSY-breaking sector
as well as the heavy fields in the PQ sector,
the low energy effective action includes the K\"ahler
potential and the superpotential of $A$,
\bea
K_{\rm eff}&=&f_{PQ}^2\{e^{(A+A^{\dagger})/f_{PQ}}
+e^{-(A+A^{\dagger})/f_{PQ}}\}+\Delta K_{\rm eff}
\nonumber \\
W_{\rm eff}&=&\mu_0e^{2A/f_{PQ}}H_1H_2+
\mu^{\prime}_ie^{3A/f_{PQ}}L_iH_2
\nonumber \\ 
&& +e^{A/f_{PQ}}(\lambda_{ijk}L_iL_jE^c_k
+\lambda^{\prime}_{ijk}L_iQ_jD^c_k),
\eea
where
$\Delta K_{\rm eff}$ is $A$-dependent loop corrections
involving the SUSY-breaking effects and
\bea 
\label{Frogatt}
&& \mu_0=y_0f_{PQ}^2/M_*,
\quad \mu^{\prime}_i=y^{\prime}_if_{PQ}^3/M_*^2,
\nonumber \\
&& \lambda_{ijk}=\gamma_{ijk}f_{PQ}/M_*,
\quad 
\lambda^{\prime}_{ijk}=\gamma^{\prime}_{ijk}f_{PQ}/M_*.
\eea
This shows that in our framework
the weak-scale value of $\mu_0$ and also the smallness of
$R$-parity violating couplings 
can be understood by means of the Frogatt-Nielsen mechanism 
of $U(1)_{PQ}$ with $f_{PQ}\ll M_*$ \cite{nielson}.
Although not written explicitly,
the coefficients of $B$-violating operators
$U^c_iD^c_jD^c_k$ can be easily arranged 
to be small enough to avoid too rapid proton decay,
{\it including} the decays into axino or gravitino
\cite{pdecay}.
The best lower bound on $f_{PQ}$ is from astrophysical arguments
implying $f_{PQ}\gtrsim 10^9$ GeV \cite{pqbound}.
To accommodate the LSND data, we need 
the axino-neutrino mixing mass of order 0.1 eV.
It turns out that this value   
is difficult to be obtained for $f_{PQ} > 10^{10}$ GeV.
We thus assume $f_{PQ}=10^9-10^{10}$  GeV
with $M_*=M_{GUT}$
for which $\mu_0$ takes an weak scale value
(with appropriate value of $y_0$)
and all $R$-parity violating couplings are appropriately
suppressed.

%Let us consider 
%the saxion effective potential and the axino mass.
An important issue here is the saxion stabilization.
One dominant contribution to the saxion effective potential
comes from
$\Delta K_{\rm eff}$
which is induced mainly by
the threshold effects of $T,T^c$ having the $A$-dependent mass 
$M_T
=\kappa f_{PQ}e^{A/f_{PQ}}$.
If $M_T\lesssim \lambda X$, one finds
\beq
\Delta K_{\rm eff}
\simeq  
\frac{N_T}{16\pi^2}
\frac{M_TM_T^{\dagger}}{{\cal Z}_T{\cal Z}_{T^c}}
\ln\left(\frac{\Lambda^2{\cal Z}_T{\cal Z}_{T^c}}{
M_TM_T^{\dagger}}\right),
\eeq
where $N_T$ is the number of superfields in $T$,
${\cal Z}_T$ is the K\"ahler metric of $T$, and $\Lambda$
is a cutoff scale which is of order $M_X$.
If $M_T\gtrsim \lambda X$, $\Delta K_{\rm eff}$
would be enhanced by $|\lambda X/M_T|^2$, however then
the resulting saxion effective potential can not stabilize
$\phi$ at the desired VEV \cite{cch}. 
We thus assume $M_T\lesssim \lambda X$.
%, so $\Delta K_{\rm eff}$ is given as above.
With ${\cal Z}_T|_{\theta^2\bar{\theta}^2}\approx
-m_{\rm soft}^2$, the above $\Delta K_{\rm eff}$
gives a negative-definite saxion potential
\beq
V^{(1)}_\phi\approx
-\frac{N_T}{16\pi^2}m_{\rm soft}^2|\kappa f_{PQ}|^2 e^{2\phi/f_{PQ}}. 
\eeq
There is another (positive-definite) potential from  the
$A$-dependent $\mu$-parameter:
\beq
V^{(2)}_\phi\approx e^{4\phi/f_{PQ}}
|\mu_0|^2(|H_1|^2+|H_2|^2).
\eeq
If $\kappa f_{PQ}$ is of order few TeVs,
$\phi$ is stabilized by
$V^{(1)}_{\phi}+V^{(2)}_{\phi}$ at
the desired value $e^{\phi/f_{PQ}}\approx 1$.
This requires a rather small Yukawa coupling 
$\kappa\sim 10^{-6}$ which may be a consequence of
some flavor symmetry.
For $e^{\phi/f_{PQ}}
\approx 1$, the saxion and axino masses are estimated
to be 
\bea
&& m^2_\phi=(10 - 10^2 \, {\rm keV})^2 +
\Delta m_\phi^2,  \nonumber \\
&& m_{\tilde{a}}=
(10^{-4}- 10^{-2} \, {\rm eV})+\Delta m_{\tilde{a}},
\eea
where the numbers represent the gauge-mediated contributions
for $f_{PQ}=10^9 - 10^{10}$ GeV, and
$\Delta m_\phi$ and $\Delta m_{\tilde{a}}$
are the supergravity-mediated contributions
which are generically of order $m_{3/2}$ \cite{lukas}.
In gauge-mediated SUSY breaking models \cite{gmsb}, 
the precise value of $m_{3/2}$ depends on the details of 
SUSY breaking sector. However most of models give
$m_{3/2}\gtrsim 1$ eV, implying
$m_{\tilde{a}}$ is dominated by
supergravity contribution.
In this paper, we assume 
\beq \label{mss}
 m_{\tilde{a}} \approx \Delta m_{\tilde{a}}\sim 1 \; {\rm eV}
\eeq
which would allow the axino
to be a sterile neutrino for the LSND data.
We note that although it is generically of order $m_{3/2}$,
$\Delta m_{\tilde{a}}$ can be significantly smaller
than $m_{3/2}$ when the supergravity K\"ahler potential takes 
a particular form, e.g. the no-scale form \cite{lukas}.

Having defined our supersymmetric axion model,
we discuss the full $4\times 4$ axino-neutrino mass matrix:
\begin{equation} \label{Lmass}
\frac{1}{2}m_{AB} \nu_A \nu_B
\end{equation}
where $A,B=s,e,\mu,\tau$ and $\nu_s\equiv \tilde{a}$
with $m_{ss}=m_{\tilde{a}}$.
We will work in the field basis in which
$\mu^{\prime}_iL_iH_2$ 
$(i=e,\mu,\tau)$ in $W_{\rm eff}$ are {\it rotated away}
by an appropriate unitary rotation of $H_1$ and $L_i$.
It is straightforward to see that the  axino-neutrino
mixing mass is given by
\beq \label{msi}
m_{is}=\frac{\epsilon_i\mu_0\langle H_2\rangle}{f_{PQ}}
 \approx 0.1 \left(\frac{\epsilon_i}{10^{-5}}\right)
  \left(\frac{\mu_0}{600 \,{\rm GeV}}\right)
\left(\frac{10^{10} \, {\rm GeV}}{f_{PQ}}\right) \, {\rm eV},
\eeq
where $\epsilon_i=\mu^{\prime}_i/\mu_0$ 
for $\mu^{\prime}_i$ in $W_{\rm eff}$ 
{\it before} $\mu^{\prime}_iL_iH_2$ are rotated away.
Note that this axino-neutrino mixing {\it survives} under
the unitary rotation eliminating $\mu^{\prime}_iL_iH_2$
since $L_i$ and $H_1$ have {\it different} $U(1)_{PQ}$-charges.
This  charge difference is also
responsible for the suppression of $R$-parity violating couplings
as well as the weak-scale value of $\mu_0$.
The $3\times 3$ mass matrix of active neutrinos
is induced by $R$-parity violating parameters \cite{hall}.
At tree-level,
\beq
m_{ij}\approx \frac{g_a^2\langle \tilde{\nu}^{\dagger}_i\rangle
\langle \tilde{\nu}^{\dagger}_j\rangle}{M_a},
\eeq
where  $M_a$ denote the gaugino masses.
The sneutrino VEV's $ \langle \tilde{\nu}_i\rangle$
 are determined by
the bilinear $R$-parity violations  in the SUSY-breaking
scalar potential:
$m^2_{L_iH_1}L_iH^{\dagger}_1+B^{\prime}_iL_iH_2$.
In our model, nonzero values of $m^2_{L_iH_1}$  and $B^{\prime}_i$
at the weak scale arise through  
renormalization group evolution 
(RGE), mainly by the coupling $\lambda'_{i33}y_b$ 
where $y_b$ is the $b$-quark Yukawa coupling \cite{hwang}.
Moreover, $BH_1H_2$ arises also through RGE which
predicts a large $\tan\beta\approx 40-60$
\cite{sarid}.  
We then find
\begin{equation} \label{mij}
m_{ij}\approx 10^{-2} t^4 
 \left({\lambda'_{i33}y_b\over 10^{-6}}\right)
\left({\lambda'_{j33} y_b \over 10^{-6}}\right) 
        \, {\rm eV}  
\end{equation}
where $t=\ln(M_X/m_{\tilde{l}})/\ln(10^3)$ for
the slepton mass $m_{\tilde{l}}$.
Here we have taken
$m_{\tilde{l}} \approx 300$ GeV and $\mu_0 \approx 2 m_{\tilde{l}}$
which has been suggested to be the best parameter range for correct
electroweak symmetry breaking \cite{sarid}.  
There are also bunch of loop corrections to $m_{ij}$
from $R$-parity violating couplings \cite{kang},
however in our case they are too small to be
relevant.

Let us now see how nicely all the neutrino masses and mixing
parameters are fitted in our framework.
The analysis of Ref.~\cite{giunti} leads to 
the four parameter regions, R1--R4 of Table 1,
accommodating the LSND with short
baseline results.
In our model, Eqs.~(\ref{mss}) and (\ref{msi})
can easily produce these LSND values of
the largest mass eigenvalue
$m_4 \approx m_{ss}=m_{\tilde{a}} \sim 1$ eV
and the mixing matrix elements 
$U_{i4}\approx m_{is}/m_{ss}\approx 0.1$.
The effective $3\times 3$ mass matrix of active neutrinos
is then given by
\begin{equation} \label{meff}
 m^{\rm eff}_{ij}=m_{ij}-m_{is}m_{js}/m_{ss}.
\end{equation}
Upon ignoring the small loop corrections,
this mass matrix has rank two, and can be written as
\newcommand{\x}{{\rm x}}
\newcommand{\y}{{\rm y}}
\newcommand{\z}{{\rm z}}
\newcommand{\w}{{\rm w}}
\begin{equation} \label{ses}
 m^{\rm eff}_{ij} = m_\x \hat{\x}_i  \hat{\x}_j
    +  m_\y \hat{\y}_i  \hat{\y}_j 
\end{equation}
where $\hat{\x}_i$ and $\hat{\y}_j$ are the unit vectors 
in the direction of $m_{is}$ and $\langle \tilde{\nu}_j \rangle$,
respectively.
Remarkably,  the mass scale $m_\x\sim (m_{is}/m_{ss})^2 m_{ss} 
\sim 10^{-2}$ eV gives the 
right range of the atmospheric neutrino mass.
Eq.~(\ref{mij}) shows that $m_\y$ is also
in the range of $10^{-2}$ eV,
so $m^{\rm eff}$ would be able to provide the right 
solar neutrino mass {\it unless} $\Delta m^2_{sol} \ll 10^{-4}$ eV$^2$.
Note from Eq.~(\ref{Frogatt})
that the typical size of $\epsilon_i, \lambda_{ijk}, \lambda'_{ijk}$ 
is around $10^{-6}$  for $f_{PQ}\approx
10^{10}$ GeV and $M_*\approx 10^{16}$ GeV.
%Let us discuss how  the atmospheric and solar 
%neutrino oscillations can be obtained from   
%the mass matrix  (\ref{ses}) of three active neutrinos.
To discuss the consequeces of the mass matrix (\ref{ses}) on 
the atmospheric and solar neutrino oscillations of three
active neutrinos, we diagonalize it to find the mass eigenvalues
%Diagonalizing (\ref{ses}), we find the mass eigenvalues
\begin{equation}
 m_{2,3} = {1\over2} \left(m_\x + m_\y \pm 
             \sqrt{ (m_\x + m_\y\cos^22\xi)^2+m_\y^2\sin^22\xi} \right)  
\end{equation}
and the $3\times 3$ mixing matrix 
\begin{equation}
 U = ( \hat{\z}^T , 
       \hat{\w}^T c_\theta - \hat{\x}^T s_\theta ,
       \hat{\w}^T s_\theta + \hat{\x}^T c_\theta)  \,.
\end{equation}
where $\hat{\z}\equiv \hat{\x}\times\hat{\y}/|\hat{\x}\times\hat{\y}|$,
$\hat{\w}\equiv \hat{\x}\times \hat{\z}/|\hat{\x}\times
\hat{\z}|$,
and  $\tan2\theta= m_\y \sin2\xi/(m_\x+m_\y\cos2\xi)$
for $\cos\xi = \hat{\x}\cdot\hat{\y}$.
The Super-Kamiokande data \cite{sk-atm} combined with the CHOOZ 
result \cite{chooz} imply that $U_{\mu3}^2\approx U_{\tau3}^2
\approx 1/2$ and $U_{e3}^2\ll1$.  
The solutions to the solar neutrino problem can have either a 
large mixing angle (LA): $U_{e1}^2\approx U_{e2}^2 \approx 1/2$, or
a small mixing angle (SA): $U_{e1}^2 \approx 1$. 
This specify the first column $\hat{\z}^T$ of $U$ as
 $\hat{\z} \approx (1/\sqrt{2}, -1/2, 1/2)$ for LA
and $\hat{\z}\approx (1, 0, 0)$ for SA up to sign ambiguities.  
Since $\hat{\x}\cdot\hat{\z}=0$,
the pattern $\hat{\z}\approx(1,0,0)$ implies $\hat{\x}_e \approx 0$.
This leads to a too small $U_{e4}\approx m_{es}/m_{ss}\lesssim
10^{-2}$, so the SA solution is {\it not allowed} within our model.
Among various LA solutions to the
solar neutrino problem, {\it only} the large-angle MSW solution
with $\Delta m^2_{sol} \sim 10^{-4}\,{\rm eV}^2$ 
can be naturally fitted  since
$m_\x \approx m_\y \sim 10^{-2}$ eV in our scheme.
It is remarkable  that 
$f_{PQ}\approx 10^{10}$ GeV and $M_*\approx M_{GUT}$
lead to the right size of $R$-parity violation
yielding the desired values of $m_{is}$ and $m_{ij}$
also for the atmospheric and solar neutrino masses.

To see the feasibility of our whole scheme,
we scanned our parameter space
to reproduce the allowed LSND islands R1--R4 of Table 1
together with the following range of
atmospheric and solar neutrino parameters:
$\Delta m^2_{31}=(1-8)\times10^{-3}\,{\rm eV}^2,
\Delta m^2_{21}=(0.1-8)\times10^{-4}\,{\rm eV}^2$
and
$\tan^2\theta_{23}= 0.33-3.8$,
$\tan^2\theta_{12}=0.2-3.0$,
$\tan^2\theta_{13}\lesssim 0.055$ \cite{glb}.
Our parameter space consists of 
$m_{ss},m_{is},\lambda^{\prime}_{i33}y_b$ whose values
are centered around 1 eV, $0.1$ eV, $10^{-6}$,
respectively.
For R1 and R4, we could find some limited parameter space,
however  they need a strong alignment between 
$\hat{\x}$ and $\hat{\y}$ and 
also a large cancellation between $m_{\x}$ and $m_{\y}$.
On the other hand,  R2 and R3
do not require any severe fine tuning of parameters,
although the bi-maximal mixing is obtained by some accident.
We provide in Fig.~1 the scatter plots on the planes of
$(m_{\y},m_{\x})$ and $(\cos\xi, m_{\x})$ for the LSND island
R2. The scatter plots for R3 have similar shape.
The full details of the analysis will be presented elsewhere
\cite{cch}.

To conclude, we have shown that the 3+1 scheme of four-neutrino
oscillation can be nicely obtained in supersymmetric 
model endowed with the PQ solution to the strong CP problem
and supersymmetry breaking mediated by gauge interactions.

{\bf Acknowledgement}:
This work is  supported by BK21 project of the
Ministry of Education, KOSEF through the CHEP of KNU
and KRF Grant No.~2000-015-DP0080.

\begin{table}
\label{lsnddata}
\begin{center}
\begin{tabular}{|c|c|c|c|}
\hline
& $\Delta m_{41}^2({\rm eV}^2)$ & $|U_{e4}|$ & $|U_{\mu 4}|$ \\ \hline
R1 &  0.21-0.28 & ~0.077-0.1~ & ~0.56-0.74~ \\ \hline
R2 &  0.88-1.1 & ~0.11-0.13~ & ~0.15-0.2~ \\ \hline
R3 & 1.5-2.1 & ~0.11-0.16~ & ~0.09-0.14~ \\ \hline
R4 & 5.5-7.3 & ~0.13-0.16~ & ~0.12-0.16~ \\
\hline
\end{tabular}
\caption{Allowed regions for the LSND oscillation.}
\end{center}
\end{table}
\begin{figure}
\begin{center}
\epsfig{file=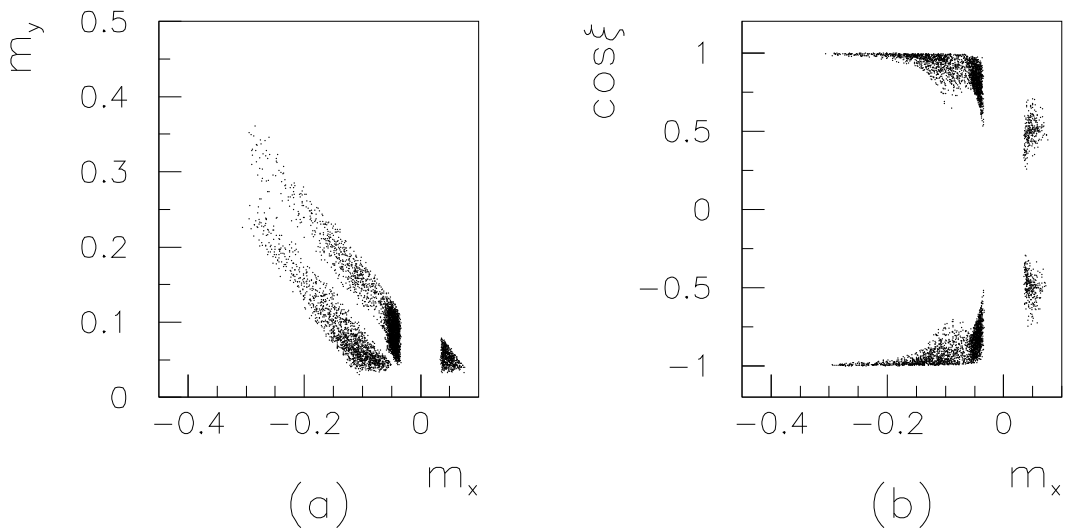, bbllx=0, bblly=330, bburx=460, bbury=567}
\caption{Scatter plots on 
(a)\,$m_\y$ vs. $m_\x$, (b)\,$\cos\xi$ vs. $m_\x$ 
reproducing the correct oscillation parameters for R2.}
\end{center}
\end{figure}

%

%\end{multicols}

\end{document}